\documentclass[twocolumn,asmfonts,amssymb,amsmath,fleqn,floatfix]{revtex4}

\usepackage[T1]{fontenc}
\usepackage{fixmath}
\usepackage[Symbol]{upgreek}
\usepackage[upmu]{gensymb}
\usepackage{tensor}
  \usepackage[dvips,final]{graphicx}

\newcommand{\abs}[1]{\ensuremath{\left\lvert #1 \right\lvert}}

\newcommand{\dbar}{{\mathchar'26\mkern-11mu\mathrm{d}}}

\newcommand{\xb}{\boldsymbol{x}}

\newcommand{\cs}{\mathrm{c}}
\newcommand{\ds}{\mathrm{d}}

\newcommand{\gs}{\mathrm{g}}

\newcommand{\ms}{\mathrm{m}}

\newcommand{\ssl}{\mathrm{s}}

\newcommand{\Lcal}{\ensuremath{\mathcal{L}}}

\newcommand{\Ocal}{\ensuremath{\mathcal{O}}}
\newcommand{\Scal}{\ensuremath{\mathcal{S}}}

\newcommand{\intfb}[1]{\ensuremath{\int\dbar^4k \,}}
\newcommand{\inttb}[1]{\ensuremath{\int\dbar^3k \,}}

\newcommand{\dbd}[2]{\ensuremath{\frac{\ds #1}{\ds #2}}}
\newcommand{\pdbd}[2]{\ensuremath{\frac{\partial #1}{\partial #2}}}

\newcommand{\ix}[1]{\indices{#1}}

\newlength\rringshift 
\newcommand*\mathrring[1]{%
   \setbox0=\hbox{$#1$}%
   \dimen0 \wd0
   \advance\dimen0 -\rringshift
   \wd0 \dimen0
   \mathring{\copy0}%
   \kern-\wd0
   \kern \rringshift
   \mathring{%
     \phantom{\copy0}%
   }%
}

\makeatletter
\def\asterisks#1{\expandafter\@asterisks\csname c@#1\endcsname}
\def\@asterisks#1{%
  \ifcase#1\or *\or **\or ***\or ****\or *****\or ******\or *******\else
  \@ctrerr\fi}

\makeatother

\begin{document}


\pagestyle{headings} \unitlength=1mm

\title{Near-Horizon Solution for DGP Perturbations}
\author{Ignacy Sawicki,$^{1,2}$ Yong-Seon Song,$^{1,3}$ and Wayne Hu$^{1,3}$ }
\email{sawickii@theory.uchicago.edu} \affiliation{
{}$^1$ Kavli Institute for Cosmological Physics, Enrico Fermi
Institute,  University of Chicago, Chicago IL 60637 \\
{}$^2$ Department of Physics,  University of Chicago, Chicago IL 60637 \\
{}$^3$ Department of Astronomy \& Astrophysics,  University of Chicago, Chicago IL 60637}
\author{}
\author{}
\affiliation{Kavli Institute for Cosmological Physics and Department
of Astronomy \& Astrophysics, University of Chicago} \date{\today}
\begin{abstract}
We develop a scaling ansatz for the master equation in Dvali,
Gabadadze, Porrati cosmologies, which allows us to solve the
equations of motion for perturbations off the brane during periods
when the on-brane evolution is scale-free. This allows us to
understand the behavior of the gravitational potentials outside the
horizon at high redshifts and close to the horizon today. We confirm
that the results of Koyama and Maartens are valid at scales relevant
for observations such as galaxy-ISW correlation. At larger scales,
there is an additional suppression of the potential which reduces
the growth rate even further and would strengthen the ISW effect.
\end{abstract}

\maketitle

\section{Introduction}
That cosmic acceleration is a fact appears indubitable. Instead of
an exotic new form of dark energy driving the acceleration, it may
be caused by a modification of gravity. Precise measurements for
gravity are only available in the range of scales from a millimeter
to that of the solar system---we do not have any direct probe of
Einstein gravity beyond these boundaries. Cosmic acceleration may
originate in a breakdown of Einstein gravity at distances beyond the
range above.

Dvali, Gabadadze and Porrati (DGP) \cite{dvali00} have proposed a
braneworld theory in which our universe is a (3+1)-dimensional brane
embedded in an infinite Minkowski bulk. Gravity propagates
everywhere, but, on the brane, an additional (3+1)-dimensional
gravitational interaction is induced. This allows for gravitational
potentials on the brane of a (3+1)-dimensional form at small
distances to evolve into (4+1)-dimensional form beyond a crossover
scale determined by the unknown energy scale for the bulk gravity.
The cosmological solution of this theory was shown to exhibit
accelerated cosmic expansion without the aid of an exotic energy
component like dark energy \cite{deffayet00}\cite{deffayet01}.

It has already been shown that the linearized field theory as defined by the
DGP model contains ghost degrees of freedom
\cite{Luty:2003vm}\cite{Nicolis:2004qq}\cite{Koyama:2005tx}\cite{Gorbunov:2005zk}\cite{Charmousis:2006pn},
or even may violate causality in certain limits \cite{Adams:2006sv}.
It is known, for instance, that the de Sitter background is unstable
to classical linear perturbations; however, it is claimed in
\cite{Deffayet:2006wp} that strong-coupling effects at small radii
around matter sources ensure that the theory remains stable.

The point of view of our work is to assume that linear
perturbation theory remains valid on the largest scales. This is motivated by the fact that the late universe is dominated by the gravitational interaction of dark-matter haloes. The internal structure of the haloes is controlled by the strongly-coupled non-linear theory. On the other hand, the radius below which strong-coupling is important for haloes is approximately equivalent to their size and therefore their interactions should be driven by the linear theory analyzed in this work.

We do find that deep into the accelerated era the spacetime
becomes unstable on the timescale of the expansion. However, this is an effect that only becomes
important far into the future and is negligible as far as the
observational impact today is concerned. We therefore assume that
during the early universe, when the theory does not exhibit
instabilities, the analysis for DGP proceeds in exactly the same way
as that for GR. Then, deep during the acceleration era,
instabilities develop and the theory may or may not be saved by
non-linear effects---an issue on which we remain agnostic. This
evolutionary history appears to be the only one which is capable of
reproducing the universe as we see it. If strong coupling effects
are important straight away and at all scales, the approximation of a homogeneous background cosmology is completely inapplicable and the DGP model would not be able to reproduce observations such as supernova
luminosities. We therefore effectively assume a best-case scenario
for DGP: should this analysis fail to predict the observations, the
model is excluded. If it passes the observational tests, a more
careful study of the effects of the strong coupling regime during
the acceleration era would be required to understand fully the
future evolution of the universe in the DGP model.

Under the above assumptions, the equations of motion for the theory of
gravity on the brane, pertinent to the study of cosmology, do not
close owing to the interaction with the bulk at first order in
perturbation theory. Koyama and Maartens \cite{koyama05} have used a
quasi-static approximation, valid well within the horizon, to
investigate structure formation at smaller scales. This solution
shows the essential role that the bulk plays in correcting the
gravitational potentials, reducing the growth rate. Also, Lue \emph{et al.} \cite{Lue:2004rj} have reached a similar conclusion using a different approach, including non-linearities in their calculations.

In the following, we present a new scaling ansatz for the master
equation, allowing us to solve the equation and calculate the
resulting cosmological evolution at all scales for high redshifts,
and close to the horizon today.  In \S \ref{s:eom}, we review the
linearized equations of motion for DGP.  We present our scaling
solution in \S \ref{s:scaling} and discuss its cosmological
implications in \S \ref{s:cosmo}.  We study the robustness of the
scaling solution in the Appendix and discuss these results in \S
\ref{s:discussion}.

\section{DGP Equations of Motion}
\label{s:eom}

\subsection{Background}

In the DGP model, gravity alone propagates in the bulk,
and the 5D gravitational theory is complemented by an induced 4D
Ricci scalar restricted to the brane. We assume that both the bulk
and the brane have zero tensions, i.e.\ the cosmological constants
are zero.
We thus start off with the basic DGP action:
\begin{equation}
    \Scal = \int\!\ds^5 x \, \sqrt{-g} \left[ \, \frac{^{(5)} R}{2\kappa^2}
    + \delta(\chi)\left(
    \frac{^{(4)} R}{2\mu^2} +\Lcal_\text{SM} \right)\right] \, ,
\end{equation}
a 3-brane embedded in an empty bulk, with all the standard-model
fields localized on the brane at $\chi=0$. The constants $\mu^2$ and
$\kappa^2$ define the energy scales of the theories of gravity: one
is Newton's constant, $\mu^2 = 8\uppi G$, the other represents the
energy scale of the bulk gravity.

As shown by Dvali \emph{et al.} \cite{dvali00}, the ratio of the two
scales defines a cross-over radius beyond which the four-dimensional
gravitational theory transitions into a five-dimensional regime
\begin{equation}
    r_\cs = \frac{\kappa^2}{2\mu^2}\,.
\end{equation}
This scale is chosen to be of the order of the current Hubble length
so that the acceleration of the expansion today results from the 4D
to 5D transition. For illustrative purposes, we will take in our
calculations $r_\cs H_0 = 1.32$, in a matter and radiation universe
with $\Omega_\ms = 0.24$ and $h = 0.66$. When comparing with
$\Uplambda$CDM, the concordance set of parameters will be used:
$\Omega_\ms = 0.25$, $h = 0.72$. The two sets of parameters
represent cosmologies which are best fits to supernova luminosity
data (SNLS \cite{Astier:2005qq}) and the distance to the
last-scattering surface (WMAP \cite{Spergel:2006hy}), but assuming
that the universe is flat. We discuss the fits to these data in a
companion paper, Song \emph{et al.} \cite{Song:2006}.

For the purpose of cosmological calculations, we would like to
reduce the five-dimensional braneworld to an effective theory on the
brane, which could be studied using the usual range of
four-dimensional tools. Using a 4+1 decomposition of the theory
\cite{shiromizu99}\cite{maeda03}\cite{sawicki05}, we can derive the
effective on-brane equations of motion---a set of modified Einstein
equations
\begin{equation}
    G\ix{_\mu_\nu} = 4r_\cs^2 f_{\mu\nu} - E_{\mu\nu} \,,
    \label{e:me}
\end{equation}
where the Greek indices range across all the dimensions, $\mu,\nu =
0,1,\ldots,4$. $f_{\mu\nu}$ is a tensor quadratic in the
4-dimensional Einstein and energy-momentum tensors,
\begin{align}
    f_{\mu\nu} &\equiv \frac{1}{12}A A_{\mu\nu} -
    \frac{1}{4}A\ix{_\mu^\alpha}A_{\nu\alpha} +
    \frac{1}{8}g_{\mu\nu} \left(A_{\alpha\beta}A^{\alpha\beta}-
    \frac{A^2}{3} \right)\,, \\
    A_{\mu\nu} &\equiv G_{\mu\nu}-\mu^2 T_{\mu\nu} \,,
\end{align}
while $E_{\mu\nu}$ is the bulk Weyl tensor projected onto the brane
using the vector normal to the brane, $n^\mu$
\begin{equation}
    E_{\mu\nu} \equiv C_{\alpha\mu\beta\nu}n^\alpha n^\beta \,.
\end{equation}
It can be shown that for branes with maximally-symmetric spatial
hypersurfaces, the projected Weyl tensor must take the form $Ca^{-4}$, with $C$ a constant (see \cite{Binetruy:1999hy}: it is effectively a constant of integration for the background). Since the Weyl tensor is traceless, it will dilute as rapidly as radiation and will become irrelevant at late times, if it doesn't completely dominate the dynamics initially. We will therefore set $C=0$, allowing us to find the modified Friedman equation for the background evolution of the cosmology on the brane. In the case of the flat brane, which is the only one to be considered here,
\begin{equation}
    H^2 \mp \frac{H}{r_\cs} = \frac{\mu^2\rho}{3}\,,
\end{equation}
leading to the result that, for the upper-sign selection, the
cosmology tends to a de-Sitter phase as the matter density gets
diluted away, potentially providing a model for the observed
acceleration \cite{deffayet:00cosmo}. This is the choice we will
make henceforth.

In addition to $r_\cs$, there is another scale present in the
theory, which thus far we have ignored: the strong-coupling scale
given by
\begin{align}
    r_* = (r_\cs^2 r_\gs)^{1/3}
\end{align}
where $r_\gs$ is the Schwarzschild radius of the point source under
consideration. Beneath this radius, the linear approximation
developed above is not valid and the theory returns to 4D GR with
small corrections. It is not understood how two sources superimpose
when the linear regime is not valid. However, the linear density
fields are actually constructed out of a spatial average of dark-
matter halos, each of which has $r_*$ comparable to its radius. Thus
we should be able to use the linear theory to discuss the
gravitational dynamics of the spatial averages of the halos, but
most likely not their internal structure.
\subsection{Linear Perturbations}
An observer restricted to the brane will perceive the universe as purely four dimensional. Therefore, the most general linear scalar perturbations to the induced flat four-dimensional metric can be parameterized by
\begin{equation}
 \ds s^2  = -(1+2\Psi)\ds t^2 + a^2(1+2\Phi)\ds \xb^2
 \, ,
\end{equation}
while he linearized energy-momentum tensor can be written down as
\begin{align}
    T\ix{^0_0} &= -\rho(1+\delta) \,,\\
    T\ix{^0_i} &= (1+w)\rho\partial_i q \,, \\
    T\ix{^i_j}  &= \rho( w+ c_{\rm s}^{2}\delta)  \delta\ix{^i_j} + w\rho\left(\partial^i\partial_j
    - \frac{1}{3}\delta\ix{^i_j}\right)\pi \,.
\end{align}
Here $\rho$ is the density of the cosmological background, $w \equiv
p/\rho$ is the background equation of state parameter, and
$c_\ssl^{2}=\delta p/\delta\rho$  is the sound speed for the
pressure perturbations.

The full 5D perturbations of the bulk are richer: Deffayet in \cite{deffayet02b} has shown that their effect on the brane can be reduced to the presence of perturbations to the Weyl tensor as an additional source of stress-energy. The precise relationship between the various components of the Weyl tensor is determined by their relationship to a master variable (see \S\ref{s:master}), which in turn encodes all the gauge-invariant perturbations to the full 5D metric throughout the bulk.  We define the linear scalar perturbations to
the Weyl tensor by
\begin{align}
E\ix{^0_0} &= -\mu^2\rho\delta_E \,, \\
E\ix{^0_i} &= \mu^2\rho\partial_i q_E \,, \\
E\ix{^i_j} &= \mu^2 w_E\rho\left[ \delta_E\delta\ix{^i_j}+
\left(\partial^i\partial_j-\frac{1}{3} \delta\ix{^i_j}  \partial^2\right)\pi_E\right]
\,.
\end{align}
 The Weyl
tensor is traceless, and hence the pressure perturbation will behave like
radiation, i.e.\ $w_E = 1/3$. The linearized Poisson and anisotropy
equations then are
\begin{align} \frac{k^2}{a^2}\Phi &= \frac{\mu^2\rho}{2}
\frac{2Hr_\cs}{2Hr_\cs-1}\Delta
+\frac{\mu^2\rho}{2}\frac{1}{2Hr_\cs-1}\Delta_E \,,
\label{e:poisson}
\end{align}
\begin{align}
\Phi+\Psi &= -\left[1+ \frac{1}{2r_\cs H
\left( 1+\frac{\dot{H}}{2H^2}\right) -1}\right] \mu^2\rho a^2 w
\pi \notag\\
&\qquad - \frac{1}{2r_\cs H \left(1+\frac{\dot{H}}{2H^2}\right) -1}
\frac{\mu^2 \rho}{3}a^2 \pi_E \,, \label{e:aniso}
\end{align}
with $\Delta$ the comoving density contrast and the equivalent
definition for $\Delta_{E}$:
\begin{align}
    \Delta = \delta - 3H(1+w)q\,, \quad \Delta_E = \delta_E -
    3Hq_E\,.
\end{align}

The equations of motion for the energy momentum
tensor are given by the conservation law $\nabla^{\mu}T_{\mu\nu} =0$ supplemented
by equations of state that define the stress fluctuations
\begin{align}
&\dbd{}{t} \left[ {\delta \over 1+w } \right] - {k^2 \over a^2} q = -3 \dot\Phi\,, \\
&\dot q - 3 c_\ssl^2 q H +  {c_\ssl^2}{\delta \over 1+w} -{2\over
3}{w \over 1+w} k^2\pi = -\Psi \,.
\end{align}
We have here assumed adiabatic pressure fluctuations  in the multicomponent matter system $c_\ssl^2 = \dot p /\dot \rho$.

On the other hand, the Weyl tensor is not separately conserved and
its equations of motion come from the Bianchi identity,
$\nabla^{\mu}G_{\mu\nu}=0$, as applied to Eq.~\eqref{e:me},
\begin{align}
    \nabla^\mu E_{\mu\nu} = 4r_\cs^2 \nabla^\mu f_{\mu\nu} \,.
\end{align}
We can rewrite this as
\begin{align}
    &\dot{\delta_E} + (1-3w)H\delta_E - \frac{k^2}{a^2}q_E = 0\,, \\
    &\dot{q_E} - 3w H q_E +
    \frac{1}{3}\delta_E - \frac{2}{9} k^2 \pi_E = S \,,
\end{align}
where the source term
\begin{align}
S\equiv
\frac{2r_\cs\dot{H}}{3H}\left[ \frac{\Delta+\Delta_E}{1-2Hr_\cs} +
\frac{k^2(w\pi +
\pi_E/3)}{1-2Hr_\cs\left(1+\frac{\dot{H}}{2H^2}\right)} \right] \,.
\end{align}
Thus a non-zero Weyl tensor is unavoidably generated by matter
perturbations in linear theory  \cite{koyama05}.

In order to close the above equations, we need the analogue of an equation
of state to relate the Weyl anisotropic stress $\pi_{E}$ to the other components of the Weyl tensor
 $\delta_E$ and $q_E$.   Unlike the relationship between the Weyl pressure and energy
 density, this relation requires a consideration of perturbations in the bulk.

\subsection{Master Equation}
\label{s:master} In \cite{mukohyama00}, Mukohyama showed that for maximally-symmetric five-dimensional space-times, the full five-dimensional linear scalar perturbations in the bulk can be described using a master variable $\Omega$. Since the bulk being considered is just Minkowski, and the brane is assumed flat, we can parameterize the unperturbed background 5D metric by
\cite{deffayet00b}
\begin{equation}
    \ds s^2 = -n(y,t)^2\ds t^2 + b(y,t)^2\ds\xb^2 + \ds y^2\,,
    \label{e:5ds}
\end{equation}
where the brane sits at $y=0$ and
\begin{equation}
    b = a(1+H\abs{y})\,, \quad n = 1 +
    \left(\frac{\dot{H}}{H}+H\right)\abs{y}\,.
\end{equation}
This parameterization is also valid for branes with non-zero curvature
provided that $\abs{\Omega_k} \ll 1$. The master variable then obeys
a hyperbolic equation of motion:
\begin{equation}
-\left(\frac{\dot\Omega}{nb^3}\right)^. +
\pdbd{}{y}\left(\frac{n}{b^3}\pdbd{\Omega}{y}\right) -
\frac{nk^2}{b^5}\Omega = 0 \,. \label{e:master}
\end{equation}
We can then express all the gauge-invariant perturbations to the 5D bulk as functions of derivatives of the master variable. In particular, Deffayet
\cite{deffayet02b} has shown that the components of the Weyl tensor evaluated on the brane can be expressed as
\begin{align}
&\mu^2\rho\delta_E = -\frac{k^4\Omega}{3a^5} \Big|_{y=0} \,,
\label{e:deltaeomega}\\
&\mu^2\rho q_E = -\frac{k^2}{3a^3}\left(\dot\Omega - H\Omega\right) \Big|_{y=0} \,, \\
&\mu^2\rho\pi_E =
-\left.\frac{1}{2a^3}\left(3\ddot{\Omega}-3H\dot\Omega +
\frac{k^2}{a^2}\Omega - \frac{3\dot{H}}{H}\pdbd{\Omega}{y}\right)
\right|_{y=0} \,. \label{e:pieomega}
\end{align}
We will hereafter implicitly assume evaluation at $y=0$ for the
master variable in the on-brane equations where no confusion might
arise. We can now rewrite the Bianchi identity in terms of the
master variable, obtaining, after assuming that the cosmological
fluid has no anisotropy stress,
\begin{align}
\ddot\Omega & - 3HF(H)\dot\Omega \notag\\
&\qquad +
\left( F(H)\frac{k^2}{a^2} +
\frac{H}{K(H)r_\cs} +
\frac{2Hr_\cs-1}{r_\cs} R H
 \right)\Omega \notag\\
&\qquad  =
\frac{2a^3}{k^2K(H)}\mu^2\rho\Delta\,, \label{e:bianchi}
\end{align}
where $R$ expresses the derivative across the brane
\begin{equation}
    R\equiv
    \frac{1}{H\Omega} \pdbd{\Omega}{y} \Big|_{y=0}
        \label{e:R} \,.
\end{equation}
We have defined two new functions of the Hubble parameter
\begin{align}
F(H) \equiv
\frac{2Hr_\cs\left(1+\frac{\dot{H}}{3H^2}\right)-1}{2Hr_\cs-1}\,, \\
K(H) \equiv
\frac{2Hr_\cs-1}{2Hr_\cs\left(1+\frac{\dot{H}}{2H^2}\right)-1}\,.
\end{align}
Were it not for the $\partial \Omega/\partial y$ derivative across the brane in $R$,
this equation would be a simple dynamic equation for $\Omega$,
which, given an evolution equation for the source $\Delta$, could be
solved as a coupled equation.  The role of the master equation is to
define $R$, the relationship between $\partial \Omega/\partial y$ and $\Omega$.

Koyama and Maartens \cite{koyama05} adopted a quasi-static approach
to solve these equations.  The master equation then implies that
the gradient
\begin{equation}
R = -{ k \over a H}\,.
\end{equation}
In the Bianchi identity the time-derivative and brane-derivative
terms are neglected compared to those of order $(k/aH)^2\Omega$.
This quasi-static approach
leads to their solution to which we will refer henceforth as
``QS''
\begin{align}
\Omega_{\rm QS}= \frac{2 a^5}{ k^4 F(H) K(H) }\mu^2\rho\Delta \,.
\label{e:KMOmega}
\end{align}
This solution is equivalent to a
closure relationship for $\pi_{E}$ in terms of $\delta_{E}$ through
Eqs.~(\ref{e:deltaeomega}) and (\ref{e:pieomega}). Using this, we can define the quasi-static limit (QS) of DGP gravity where the Poisson and anisotropy equations become
\begin{align}
    \frac{k^2}{a^2}(\Phi-\Psi) &= \frac{\mu^2\rho}{2}\Delta \\
    \frac{k^2}{a^2}(\Phi+\Psi) &=  -\frac{\mu^2\rho}{6\beta}\Delta \,,
\end{align}
with
\begin{equation}
    \beta \equiv 1 -2r_\cs H\left(1 + \frac{\dot{H}}{3H^2}\right) \,.
\end{equation}
These equations are equivalent to the linear limit of the results obtained by Lue \emph{et al.} \cite{Lue:2004rj}.

In the next two sections, we will show how the quasi-static solution is
dynamically achieved for the perturbations shortly after horizon
crossing and discuss large-scale deviations from this solution.

\section{Scaling Solution to Master Equation}
\label{s:scaling}

\subsection{Causal Horizon}
The master equation is a wave equation sourced by the comoving
density perturbations on the brane through the Bianchi identity
\eqref{e:bianchi}.

Since, in appropriate coordinates, the bulk is just Minkowski, the
evolution of $\Omega$ in the bulk can be seen as a normal propagating wave
given a boundary condition from
the behavior on the brane.   Beyond the causal horizon, the
bulk should remain unperturbed.
This causal
horizon must be invariant in all coordinizations of the bulk;
therefore we can locate it by finding the null geodesic of
Eq.~\eqref{e:5ds}, giving us the $y$-position of the horizon as
\begin{align}
    \xi = y_\text{hor}H = aH^2\int_0^{a} \frac{\ds \tilde a}{\tilde a^2H(\tilde a)^2}
    \,. \label{e:xi}
\end{align}
Before the acceleration epoch, this reduces to $\xi=1/(2+3w)$ for a
cosmology with a constant equation of state parameter, i.e.\ $\xi =
1/3$ for a radiation-only cosmology and $\xi = 1/2$ for a
matter-only cosmology. We are making the assumption that the
universe has not gone through a period of inflation, which would
have moved the horizon much further out. If inflation did take
place, it is not unreasonable to expect that any perturbations that
existed prior to the inflationary phase will have been pushed far
away and the bulk will start in an unperturbed state in the vicinity
of the brane at the beginning of radiation domination, resulting in
a causal horizon equivalent to that of the cosmology with no
inflation.

The constancy of the horizon during the domination of a particular
fluid suggests that we can define a new variable, in which the
horizon will remain fixed at all times, lying at $x=1$
\begin{equation}
    x\equiv \frac{yH}{\xi} \,.
\end{equation}
We can then recast the master equation in $x$ and solve it as a
boundary value problem with the value of $\Omega$ at the horizon,
$x=1$, set to zero.

\subsection{Scaling Ansatz}

The second boundary condition needed to solve the master equation is
the behavior of the master variable on the brane. During epochs when
the source remains scale-free and the Bianchi dynamics also do not
change, one would expect that the master variable also obeys a
scaling ansatz on the brane $ \Omega |_{y=0} = A a^{p}   \,, $ where
$A$ and $p$ are constants.  Likewise, during such epochs we expect
the master variable in the bulk to reach a stable solution in the
variable $x$, the distance in units of the causal horizon, for a
given wavenumber $k/aH$.

We therefore propose a new ansatz for the solution to the master
equation \eqref{e:master}:
\begin{equation}
    \Omega = A(p) a^p G(x) \label{e:ans}\,.
\end{equation}
With this assumption, the master equation \eqref{e:master} becomes
the ordinary differential equation

\begin{widetext}
\begin{align}
&\dbd{^2G}{x^2} + \left( \frac{4h^2+h(2p-1)+2h'} {2h(1+x\xi(1+2h))}
- \frac{1+2p}{2(1+x\xi)} - \frac{h+h^2+h'}{h(1+x\xi(1+h))}
\right)\xi\dbd{G}{x} +
 \left( \frac{hp(2p-5)-2h'p}{4h^2(1+x\xi)} +
\frac{3p}{2(1+x\xi)^2} -\right. \notag\\
&\qquad - \left. \frac{(1+2h) (h(4h+2p-1)+2h')p}{4h^2(1+x\xi(1+2h))}
+ \frac{(1+h)(h+h^2+h')p}{h^2(1+(1+h)x\xi)}
 - \left(\frac{k}{aH}\right)^2
\frac{(1+(1+h)x\xi)^2}{(1+x\xi)^3(1+x\xi(1+2h))} \right)\xi^2G = 0\,,
\label{e:ODE}
\end{align}
\end{widetext}
where the primes denote differentiation with respect to $\ln
a$---the new time coordinate which will be used henceforth---and
$h\equiv H'/H$. In deriving the above, we have neglected time
derivatives of $p$ and $\xi$: the scaling ansatz is not expected to
be valid when $p$ is not a constant, i.e.\ during times when the
cosmology is undergoing a change from the domination of one fluid to
another. In addition, strictly speaking, $G$ is
actually a function of both $x$ and $k/aH$, even in the scaling
limit. We have therefore also assumed that $k/aH$ is a constant,
which is valid in the $k=0$ limit. As we explain later, as $k/aH$
approaches unity, where its time-derivatives might impact the
solution significantly, the character of Eq.~\eqref{e:ODE} changes
and terms which do not involve the derivatives of $k/aH$ dominate.

Note that, provided $w$ is constant, one of the denominators in
Eq.~\eqref{e:ODE} can be re-expressed as
\begin{align}
    1+x\xi(1+2h) = 1-x \,.
\end{align}
Thus the equation has a regular singular point $x=1$, exactly at the
causal horizon. This is not a coordinate singularity (all the
entries of the metric are regular there), but is a reflection of the
junction between perturbed and unperturbed space-times.

Supplying $H$, $p$ as a function of the scale factor is enough to
solve this ODE as a boundary-value problem, requiring that $G(x)$ be
1 on the brane and 0 at $x = 1$. This in turn gives the value of $R$
as
\begin{equation}
    R =\frac{1}{\xi}\frac{\left.\dbd{G}{x}\right|_{x=0}}{\left.G\right|_{x=0}} \,,
\end{equation}
and closes the evolution equations for the perturbations on the brane.

We will henceforth refer to this solution as the dynamical scaling
or just scaling solution and use the acronym ``DS''.

\subsection{Iterative Solution}

In practice, one does not know the scaling index $p$ a priori and,
moreover, it can change during the evolution of a $k$ mode as the
master variable leaves one scaling regime and enters another.  We
therefore solve for $p$ iteratively by demanding consistency with
the Bianchi identity.

To determine the zeroth-order solution for $p$, we substitute
 the ansatz Eq.~\eqref{e:ans} into
Eq.~\eqref{e:bianchi} to obtain
\begin{align}
A(p) &=
\frac{2a^{3-p}}{k^2K(H)H^2 [J(H) + F(H) (k/aH)^{2}}]\mu^2\rho\Delta \,,
 \label{e:peq}\\
J(H) &= p[ p + {{h}}-3F(H)]+
\frac{1}{K(H)Hr_\cs} +\frac{2Hr_\cs-1}{Hr_\cs} R \,. \notag
\end{align}
Before the acceleration epoch, when the expansion is dominated by a
single fluid, $J(H)$ and $F(H)$ are constant, while $K(H)$ is a
simple power law in $a$. We therefore set
 $p = p^{(0)}$, the zeroth order solution for $k/aH \ll 1$
\begin{align}
p^{(0)} &= 3 + \dbd{\ln  [\rho \Delta^{(0)}/ (K(H) H^{2})]}{\ln a}
\,.
\end{align}
Note that this definition allows transitions between scaling regimes
where $w$ changes.  We shall see that the solutions for $R$ become
independent of $p$ for $k/aH\gg 1$ and so we use this as the zeroth
order solution for all modes, both super- and subhorizon.

\begin{figure}[ht]\begin{centering}
\includegraphics[width=\columnwidth,height=6.9cm]{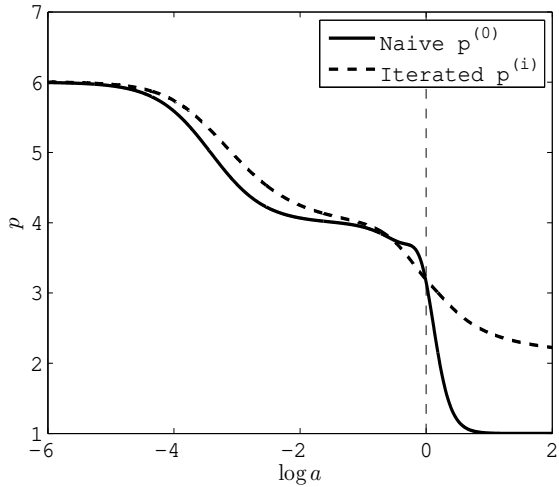}
\caption{Value of exponent $p$ in the scaling
solution $\Omega|_{y=0} \propto a^{p}$ for
superhorizon modes, plotted as a function of the scale factor.  The
$p^{(0)}$ calculation assumes that the density perturbation $\Delta$
follows Eq.~\eqref{e:growth} at all times. The $p^{(i)}$ result is
the output of an iterative process, where $p$ is used to calculate
$R$ and hence the evolution of $\Omega$ which is in turn used to
derive a correction to $p$. \label{f:p}}
\end{centering}\end{figure}

Finally we need a zeroth order solution for $\Delta^{(0)}$.
Modes of interest to large-scale cosmological tests are superhorizon
during radiation domination and enter the horizon either during
matter domination or the current acceleration epoch. Before the
acceleration epoch and in the absence of anisotropic
stress, these modes obey \cite{hu94}
\begin{align}
\Delta^{(0)} \propto \frac{D^3 +
\frac{2}{9}D^2-\frac{8}{9}D-\frac{16}{9}+\frac{16}{9}\sqrt{D+1}}{D(D+1)}\,,
\label{e:growth}
\end{align}
where $D \equiv a/a_\text{eq}$. In order to obtain the zeroth-order
solution for $p^{(0)}$ we assume that the growth prescribed by
equation Eq.~\eqref{e:growth} continues until today (see
Fig.~\ref{f:p}).

Given this, we solve the master equation Eq.~\eqref{e:ODE} to obtain
the off-brane gradient, $R$, and then dynamically solve the Bianchi
identity Eq.~\eqref{e:bianchi} coupled to the cosmology, for the
particular mode. Once the dynamic evolution for the first-order
solution $\Omega^{(1)}$ is obtained, we can iteratively improve our
estimation of $p$ by numerically calculating
\begin{equation}
p^{(i)} = \dbd{\ln \Omega^{(i)}}{\ln a}
\end{equation}
and repeating the above prescription. We find in the next section
that this procedure
converges quickly and alters the value of $p$ only when $p$ is not a
constant, as expected. We display the effects of the iteration on
$p$ in Fig.~\ref{f:p}.

\section{Cosmological Implications}
\label{s:cosmo}
\subsection{Limiting Cases and Numerical Solutions}

 The evolution of the master variable exhibits
several distinct phases that are distinguished
 by the on-brane scaling evolution: that during
radiation domination, matter domination and the de-Sitter
acceleration phase both inside and outside the
horizon. The scaling of the Bianchi identity
Eq.~\eqref{e:bianchi} determines the value of $p$ at a particular
scale factor, while the solution to the master equation
Eq.~\eqref{e:ODE} determines the off-brane gradient $R$. To better
understand the nature of the solution and how it impacts
perturbation evolution, we will compare the analytic expectations
to the full numerical results in the various phases.

\begin{figure}[ht]\begin{centering}
\includegraphics[width=\columnwidth]{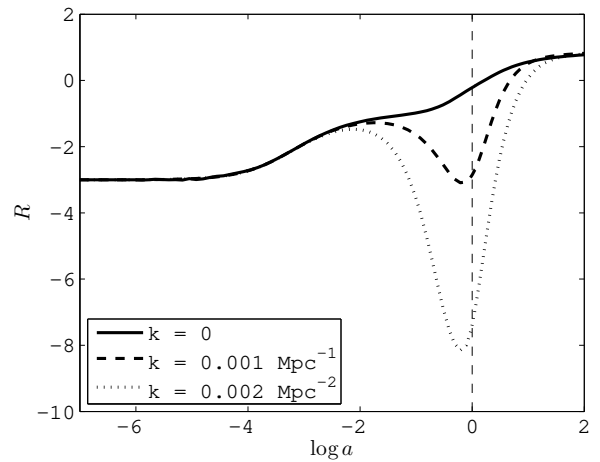} \caption{Evolution of ratio of
off-brane gradient to master variable, $R$, as defined in
Eq.~\eqref{e:R}, for a selection of modes.
On superhorizon scales, $R$ is constant whenever $p$ is a constant.
Once the mode enters the horizon
it rapidly approaches $R = -k/aH$. As the universe enters the
de-Sitter phase, the modes again leave the horizon and $R$ asymptotes to
$1$. \label{f:Rplot}}
\end{centering}\end{figure}

\subsubsection{Superhorizon Modes}
During radiation do\-mi\-nation, $Hr_\cs \gg 1$, $H'/H =
-2+a/2a_\text{eq}$ and $F(H)=1/3$. The particular combination in the
denominator of $K(H)$ causes the first-order contribution of $H'/H$
to cancel, leaving us with $K(H) \propto a^{-1}$. The Bianchi
identity Eq.~\eqref{e:bianchi} for superhorizon modes then dictates
that
\begin{equation}
    A \propto a^{4-p} \Delta \,.
\end{equation}
Since $\Delta \propto a^2$ during radiation domination, this gives
$p = 6$ for superhorizon modes. Inserted back into the master equation
under the scaling
ansatz Eq.~\eqref{e:ODE}, this value of $p$ implies
\begin{align}
 R &= -3, \qquad (k/aH \ll1, \, \text{radiation domination})\,.
\end{align}
With $R$ determined, the equations of motion for the
perturbations on the brane are closed.

This analytic expectation also serves as the initial conditions for
the numerical scaling solution.  In practice, we begin the integration at
$a = 10^{-6}$, when all modes of interest are outside the horizon.
The numerical solution for $p$ is shown in Fig.~\ref{f:p} and for
$R$ in Fig. \ref{f:Rplot}.  Note that, in the radiation-dominated era,
their values stay stable at the analytic prediction for all
iterations of the solution.

\begin{figure}[ht]\begin{centering}
\includegraphics[width=\columnwidth]{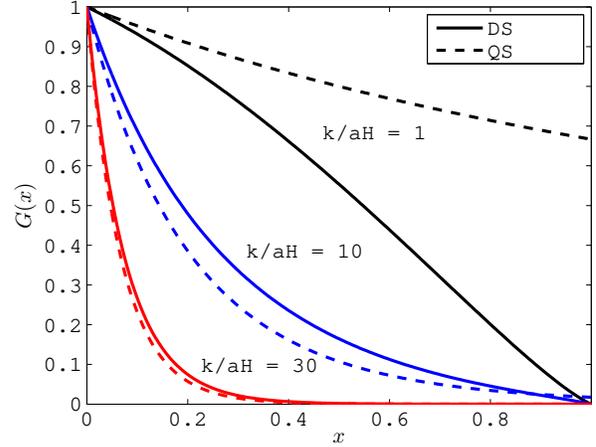}
\caption{Off-brane profile for $G(x)$ obtained by solving
equation~\eqref{e:ODE} during matter domination ($\log a = -2$),
compared to off-brane profiles for the quasi-static (QS) solution.
For high $k/aH$ the profiles
are very narrow and effectively independent of the position of the
causal horizon: they penetrate very little into the bulk and the
behavior of the solution is practically independent of the value of
$p$. In this regime, the QS solution is practically coincident with
the scaling solution. For modes with low $k/aH$, the solution is
non-zero in the whole interval $x \in [0,1)$ and therefore it
depends strongly on the value of $p$. QS severely underestimates the
gradient of the profile in this regime. \label{f:gplot}}
\end{centering}\end{figure}

The large-scale modes of interest remain outside the horizon during
the whole radiation-dominated epoch.    In the matter-dominated
epoch, the evolution outside the horizon can be obtained by noting,
$F(H)=1/2$, $K(H)=4$, $H'/H=-3/2$, and $Hr_\cs \gg 1$. The Bianchi
identity dictates that
\begin{equation}
A \propto a^{3-p} \Delta \,.
\end{equation}
Given that $\Delta \propto a$, $p=4$.
Since the matter dominated solution is of particular interest,
we explicitly give the master equation under the scaling ansatz
\begin{align}
&\dbd{^2G}{x^2} +
\left(\frac{7-2p}{4(x-1)}-\frac{1+2p}{2(x+2)}-\frac{1}{x-4}
\right)\dbd{G}{x}
+\notag\\
&\quad+ \left( \frac{p(2p-7)}{12(x-1)} +
\frac{p(5-2p)}{12(x+2)} + \frac{p}{6(x-4)} + \frac{3p}{2(x+2)^2} + \right.\notag\\
&\quad\left.+ \left(\frac{k}{aH}\right)^2
\frac{(x-4)^2}{8(x-1)(x+2)^3} \right)G =0\,. \label{e:matode}
\end{align}
This is solved as a boundary value problem with boundary conditions
$G(0)=1$ and $G(1)=0$. The form of the numerical solution to this
equation is shown in Fig.~{\ref{f:gplot}}.  In the large-scale
limit, the gradient reaches
\begin{align}
 R &= -1, \qquad (k/aH \ll1, \, \text{matter domination})\,.
\end{align}
In the numerical solution of  Figs.~\ref{f:p} and \ref{f:Rplot},
these values of $p=4$ and $R=-1$ are achieved gradually as the
expansion becomes matter dominated.  The iteration of the numerical
solution in fact further smooths the transition until a stable form
is achieved as would be expected. Note also that $R$ is very
insensitive to $k/aH$, provided it be less than 1 such that the mode
is larger than the horizon.

\subsubsection{Subhorizon Modes}

The modes of interest cross the horizon either during matter domination
or the acceleration epoch.
For large values of $k/aH$, the final term of the master equation
Eq.~\eqref{e:ODE} dominates over other parts of the coefficients of
$G(x)$ and $G'(x)$.  As evidenced in Fig.~\ref{f:gplot}, in this
regime, the solution does not penetrate very far into the bulk. We
can thus expand the master equation around $x=0$, reducing it to
\begin{equation}
    \dbd{^2G}{x^2} - \xi^2(1-2\xi x)\left(\frac{k}{aH}\right)^2G = 0 \, .
\end{equation}
This matches the quasi-static approximation up to first order in
$x$. Therefore, for $k/aH \gg 1$, the relation giving the gradient
on the brane is exactly as in the QS approximation
 \cite{koyama05}, with no
dependence on the value of $p$:
\begin{equation}
    R = -\frac{k}{aH} \qquad (k/aH \gg1, \, \text{matter/acceleration}) .
\end{equation}
The numerical solution for $R$ evolved through horizon crossing is
shown in Fig.~\ref{f:Rplot}.   It reaches this scaling shortly after
horizon crossing.

Despite this independence of $p$, the master variable does achieve a
scaling form during matter domination. The Bianchi identity
\eqref{e:bianchi} can be reduced to the QS form,
Eq.~\eqref{e:KMOmega}, and implies
\begin{equation}
A \propto a^{2-p} \Delta \,,
\end{equation}
and hence $p = 3$.

In general then, $R$ is a function of both $k/aH$ and $p(a)$ and
therefore each mode needs to be followed separately through its
evolution both outside and inside the horizon. However, only outside
the horizon does the value of the off-brane gradient actually affect
the evolution on the brane, since for high $k/aH$ it is subdominant
in the Bianchi identity Eq.~\eqref{e:bianchi}.

\subsubsection{Asymptotic de-Sitter phase}
\label{s:dS} At late times, the DGP cosmology enters the
self-accelerated de-Sitter phase. During this time all modes exit
the horizon while the causal horizon in the bulk, $\xi$, grows
rapidly toward infinity. This allows us to concentrate on the $k
\ll a H$ limit. Deep into the de-Sitter phase of the expansion,
$Hr_\cs = 1$, $K(H)=F(H)=1$. The master equation \eqref{e:master}
can now be rewritten (in the scaling approximation) as
\begin{equation}
    \pdbd{^2\Omega}{y^2} - \frac{2}{1+Hy}\pdbd{\Omega}{y} -
    \frac{p(p-3)}{(1+Hy)^2}\Omega = 0 \, .
\end{equation}
This equation has an analytic solution. Assuming that the
perturbations vanish at the causal horizon, $\Omega= 0$ at $y = \xi/H$, and that
$p \neq 3/2$, we can find the off-brane gradient:
\begin{equation}
R = \frac{1}{2}\left(3 - \abs{2p-3}\right) -
\frac{\abs{2p-3}}{(1+\xi)^{|2p-3|}-1} \, . \label{e:dSR}
\end{equation}
This can now be combined with the Bianchi identity
\begin{equation}
(p^2 - 3p +1 +R)\Omega = \frac{6H_0^2r_\cs^2\Omega_m\Delta}{k^2} \,.
\label{e:dSB}
\end{equation}
It can be shown that $\Delta$ becomes a constant during the
acceleration era.

In the limit where $r_\cs^2\Delta/\Omega \rightarrow 0$ and $\xi
\rightarrow \infty$, i.e.\ at very late times, the solution to the
above two equations combined is $R=1$ and $p = 2$ or $p=1$, where we
would expect the fastest-growing mode, $p=2$, to dominate.  We plot
the numerical solutions in Figs.~\ref{f:p} and \ref{f:Rplot}.  Note
that iteration of the solution here is crucial but $p=2$, $R=1$ is
approached at late times.

The numerical results only converge to $p=2$, $R=1$ very slowly.
This can also be understood analytically.
There is a slight correction to this result arising as a result of
$\xi$ being finite, but it is already insignificant by $\log a \sim
2$. A more significant correction to the scaling behavior arises as
a result of $\Delta$ not being zero. Its size can be estimated by
assuming that $p = 2+\epsilon$. Substituting this into both
Eq.~\eqref{e:dSR} and Eq.~\eqref{e:dSB} we obtain
\begin{equation}
\epsilon = 1 - \frac{1}{2}\sqrt{4+\frac{6H_0^2
r_\cs^2\Omega_\ms\Delta}{\Omega k^2}} \, .
\end{equation}
Indeed, as $\Omega$ grows, the correction decreases until $p=2$ is
achieved. In particular, the continued growth of $\Omega$ implies
that the anisotropy at the largest scales will continue growing as
\begin{equation}
\Phi + \Psi \propto \frac{\pi_E}{a} \propto a^{1+\epsilon} \, .
\end{equation}

\subsubsection{Superhorizon Metric Evolution}

Deriving an evolution equation for the metric highlights the fact
that the metric evolution is not sensitive to the DGP modification
until the acceleration epoch.  Even in the acceleration epoch the
metric obeys a simple equation of motion on superhorizon scales.

By appropriately rearranging the linearized Einstein equations and
the Bianchi identity,
 we can obtain a single differential equation
for the gravitational potentials
\begin{align}
&\left(\Phi'' - \Psi' - \frac{H''}{H'}\Phi' +
\left(\frac{H''}{H'}-\frac{H'}{H}\right)\Psi + \right.\notag\\
&\left.\left(\frac{k}{aH}\right)^2\left[\left(c_\ssl^2 +
\frac{H'}{3H} \frac{1}{2Hr_\cs-1}\right)\Phi+\right.\right.
\notag\\
&\left.\left. +\frac{H'}{9(1+w)H}\frac{2Hr_\cs-1}{Hr_\cs-1}
\left(\Phi + \Psi\right)\right]\right)(1+w) = \notag\\
&=\frac{H'}{H}\frac{1}{2Hr_\cs}\left[(w-c_\ssl^2)\Delta_E +
\left(\frac{1}{3}-w+\frac{H'}{3H}\right)\delta_E \right] \, .
\end{align}
In the large-scale limit, $k/aH \rightarrow 0$, $\delta_E \approx
0$; if we also assume that $c^2_\ssl = w$ or that $\Delta_{E} = {\cal O}(k/aH)^{2}\Phi$
from the Poisson equation, we obtain
\begin{equation}
    \Phi'' - \Psi' - \frac{H''}{H'}\Phi' +
\left(\frac{H''}{H'}-\frac{H'}{H}\right)\Psi = 0\label{e:bert} \, .
\end{equation}
Note that this equation for the metric perturbations is the same as
found in general relativity [e.g. \cite{HuEis99} Eq.~(50)].
Therefore, before the DGP modifications change the expansion rate or
generate anisotropic stress, the evolution of the metric is
identical to general relativity.  Furthermore, as pointed out by
 Bertschinger
\cite{Bertschinger:2006aw} in the absence of anisotropic stress, $\Psi = -\Phi$ and
this equation yields solutions that depend only on the expansion
history through $H$ even in the acceleration epoch.

In DGP gravity, however, the anisotropy at the largest scales is
never negligible and, in fact, grows at late times, as discussed in
section~\ref{s:dS}. We can rewrite Eq.~\eqref{e:bert} by defining
\begin{align}
    \Phi_+ = \frac{1}{2}(\Phi+\Psi) \qquad
    \Phi_- = \frac{1}{2}(\Phi-\Psi) \, .
\end{align}
In the de Sitter era, when $H'/H = 0$ and $H''/H' =-3$, and assuming
that the two new variables obey a scaling solution with $\Phi_+ =
A_+ a^{p_+}$ and $\Phi_- = A_- a^{p_-}$, Eq.~\eqref{e:bert} becomes
\begin{align}
    A_+a^{p_+}(p_+^2+2p_+-3) + A_-a^{p_-}(p^2_-+4p_-+3) = 0\,.
\end{align}
One would expect that $\Phi_+$ would grow with exponent $p_+ = 1$,
while $\Phi_-$ would decay away with exponent $p_- =-1$. However, because of
the need to preserve the Bianchi identity with a non-zero $\Delta$,
as discussed in section~\ref{s:dS}, the scaling solution is slightly
violated. We find that $p_+ = 1 + \epsilon$ while $p_- = 1$. This
leads to the relation
\begin{equation}
    \frac{A_-}{A_+} = -\frac{\epsilon a^\epsilon}{2} \, ,
\end{equation}
where $\epsilon \rightarrow 0$ monotonically. At late times, our
solution tends to a regime where the ratio $\Phi_-/\Phi_+
\rightarrow 0$, with $\Phi_+ \propto a$.

It is interesting to note
that in this opposite limit to that studied in \cite{Bertschinger:2006aw}
where $\Phi_{-}\ll \Phi_{+}$
Eq.~(\ref{e:bert}) also becomes closed and has solutions that depend only
on the expansion history through $H$.

\begin{figure}[ht]\begin{centering}
\includegraphics[width=\columnwidth]{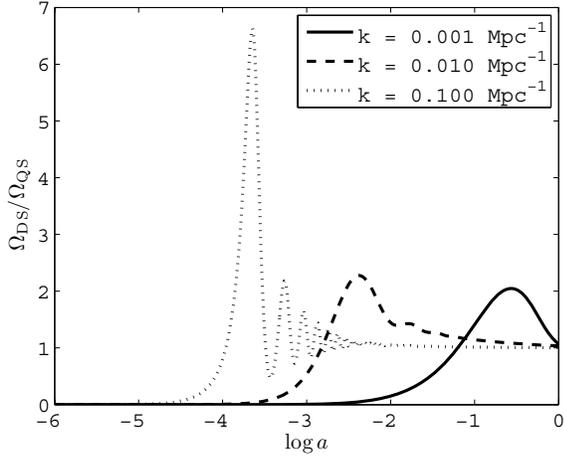}
\caption{Ratio of master variable $\Omega$ for dynamic scaling and
quasi-static solutions. In QS, $\Omega$ responds instantaneously to
changes in $\Delta$. Fully dynamic solution to the Bianchi identity
\eqref{e:bianchi} requires time to respond and eventually decays to
the QS solution. Initializing the calculation at earlier times
changes neither the scale factor at which $\Omega$ responds nor its
value today. Rapid growth occurs during the time before horizon
crossing. \label{f:Omegacomp}}
\end{centering}\end{figure}

In fact, Eq.~\eqref{e:bert} is equivalent to the statement that the Bardeen curvature, $\zeta \equiv \Phi + Hq$, is conserved during the de Sitter era, once the mode leaves the horizon. This is also true of the comoving density perturbation, $\Delta$, which saturates to a constant during the de Sitter era. However, the comoving and longitudinal hypersurfaces warp with a shift $Hq$ that grows without bound.

\subsection{Quasi-Static vs Dynamical-Scaling Solutions}

It is useful to summarize the differences between the quasi-static
(QS) and dynamical scaling (DS) solutions uncovered in the previous
section.

\begin{figure}[ht]\begin{centering}
\includegraphics[width=\columnwidth]{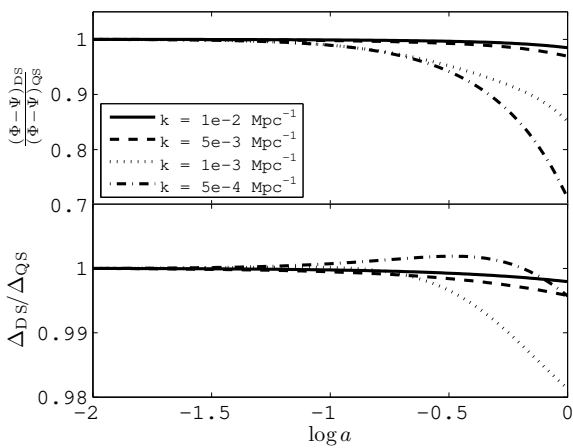}
\caption{\label{f:phicomp} Comparison of results of QS and DS
solutions: Upper panel presents the ratio of $\Phi-\Psi$ in
the two approximations. For modes with $k > 0.01$~Mpc$^{-1}$ the two
solutions differ by less than 2\%. The difference is much more
pronounced for larger scales where $\Omega$ has not decayed to the
QS value, resulting in additional decays of up to 15\%. Lower panel
presents the ratio of comoving density perturbations in the two
solutions: $\Delta$ is affected much less, with approximately a 2\%
deviation from QS at the largest scale, $k = 0.001$~Mpc$^{-1}$.}
\end{centering}\end{figure}

Beginning at the initial conditions in the radiation-dominated era,
the superhorizon value of the master variable is highly suppressed
with $\Omega_\text{DS}/\Omega_\text{QS} =
\Ocal\left((k/aH)^2\right)$ (see Fig.~\ref{f:Omegacomp}).
As the mode enters the horizon during matter domination, the DS
solution for $\Omega$ grows rapidly and then executes damped
oscillations around the QS solution.   This can be understood
analytically since the Bianchi identity takes the form of a damped
oscillator in $\Omega/a^2$ that is driven by $\Delta$. During the
time when $\Omega$ significantly deviates from the QS solution, the
Weyl corrections to the Poisson equation \eqref{e:poisson} are
suppressed, since $Hr_\cs \gg 1$, and there is no additional
correction to the gravitational potentials over and above that of
QS (see Fig.~\ref{f:phicomp}).

\begin{figure}[ht]\begin{centering}
\includegraphics[width=\columnwidth]{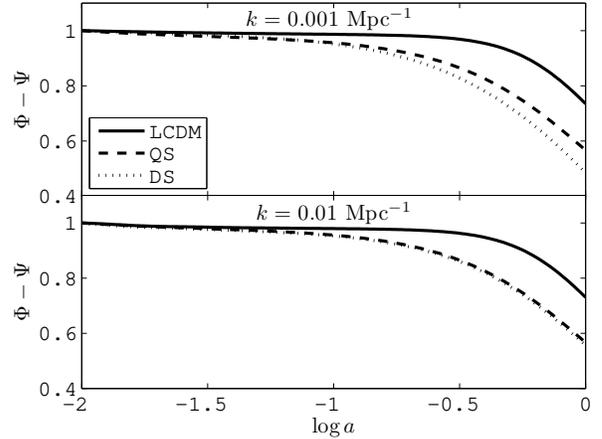}
\caption{Evolution of the principle gravitational observable $\Phi-\Psi$ for
concordance $\Uplambda$CDM, the quasistatic (QS) solution,
and our new dynamical scaling (DS) solution.
For scales $k \gtrsim 0.01$~Mpc$^{-1}$ the scaling and QS solutions
do not differ appreciably: the decay in the potentials is a little
faster than $\Uplambda$CDM as a result of slower growth of density
contrast. At larger scales, the DS solution exhibits significant additional decay owing to
the different value of $\Omega$ at late times, as
exhibited in Fig.~\ref{f:Omegacomp}. All potentials normalized to 1
at $\log a = -2$. \label{f:phi}}
\end{centering}\end{figure}

\begin{figure}[ht]\begin{centering}
\includegraphics[width=\columnwidth]{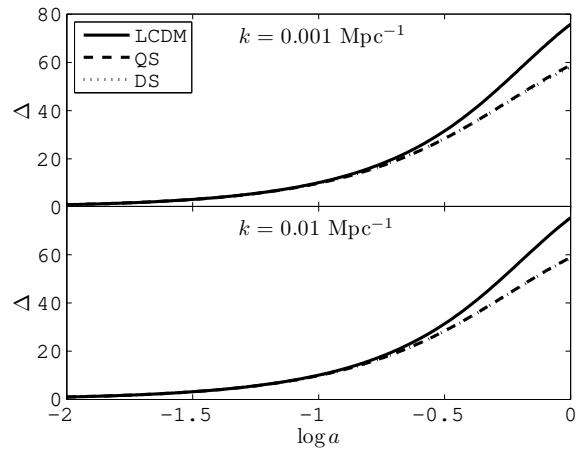}
\caption{Evolution of the comoving density contrast $\Delta$ for
concordance $\Uplambda$CDM, QS and our DS solution. The
growth function in DGP is suppressed compared to $\Uplambda$CDM,
even in the case of a flat cosmology. There is no significant difference between the scaling ansatz and the quasi-static solution, with DS departing at most by 2\% from QS at the largest scales. All quantities are normalized to 1 at $\log a = -2$. \label{f:delta}}
\end{centering}\end{figure}

We find that the results of the scaling solution match the
quasi-static results for all modes that enter the horizon well
within the matter-dominated epoch $k > 0.01$~Mpc$^{-1}$  (see
Figs.~\ref{f:Omegacomp}--\ref{f:phicomp}). For larger scales, this
is not so: as shown in Fig.~\ref{f:Omegacomp}, $\Omega$ now only
decays toward the QS solution, rather than oscillating around it.
Since at late times the Weyl perturbations are no longer suppressed,
this now makes a significant contribution to the Poisson equation,
resulting in additional decay of the potentials. As shown in
Fig.~\ref{f:phicomp}, there is a 15\% deviation from QS in
$\Phi-\Psi$ at the scale $k = 0.001$~Mpc$^{-1}$ for the chosen sets
of cosmological parameters. The
direction of this effect agrees with estimates made by Lue in
\cite{Lue:2005ya} and is such that the scaling solution is an even
worse fit to CMB anisotropy data than the quasi-static (see
\cite{Song:2006} for a discussion).

On the other hand, up to the present time, the QS solution for comoving density perturbations, $\Delta$, is a very good approximation for the DS solution at all scales. The additional suppression is of the order of 2\% for $k = 0.001$~Mpc$^{-1}$.

\section{Discussion}
\label{s:discussion} We have introduced a new scaling ansatz which
allows solutions to linear perturbations in the DGP model on all
scales less than the cross-over scale $r_\cs$ up to the present
epoch. The equations of motion for linear perturbations on the brane
require knowledge of the gradient of the so-called master variable
into the bulk. The master variable obeys a master equation in the
bulk.  To solve the master equation, it is sufficient to have two
boundary conditions, one on the brane and the other in the bulk.

Our scaling solution begins with an ansatz for the brane boundary
condition: that the evolution of the master variable is scale free
on the brane.   The second boundary condition is that the master
variable vanishes at the causal horizon in the bulk.  With these two
boundary conditions, we solve the master equation to determine the
gradient. With the gradient known, we can then replace the
scale-free ansatz with the dynamical solution and iterate the
solution until convergence.

We find that the quasi-static (QS) solution of \cite{koyama05}
is rapidly approached once the perturbation crosses the horizon.
Before horizon crossing there are strong deviations from the
quasistatic solution. For modes that crossed the horizon only
recently during the acceleration epoch, we find that the metric
perturbation $\Phi-\Psi$ decays more rapidly that the QS solution.
The QS solution itself has a stronger decay than the $\Uplambda$CDM
model.  The extra decay compared with $\Uplambda$CDM is extremely
robust to changing the gradient of the master variable into the
bulk, the one variable that is required to close the equations of
motion on the brane. We consider the observational consequences of
these results in a companion paper \cite{Song:2006}.

\begin{acknowledgments}
We thank Sean Carroll, Chris Gordon, Nemanja Kaloper,
Arthur Lue and Roman Scoccimarro for useful
conversations.
This work is supported by the U.S.\ Dept.\ of Energy contract
DE-FG02-90ER-40560. IS and WH are additionally supported at the KICP
by the NSF
grant PHY-0114422 and the David and Lucile Packard Foundation. The
KICP is an NSF Physics Frontier Center.
\end{acknowledgments}

\appendix

\section{Scaling Ansatz Robustness}
 \label{s:robust} Assuming that the scaling ansatz is
appropriate, the solution presented in this paper is correct at all
scales. However, the scale-free assumption for the solution in the bulk depends crucially on the existence of only one scale in the problem: the Hubble parameter. It is possible that through some additional physics in the vicinity of $r_\cs$ this assumption is broken and the evolution of the bulk profile will depend on both $H$ and $r_\cs$ independently.

However, any violation of scaling will enter the equations solely through the off-brane gradient $R$: this is the quantity which we obtained by
solving the master equation Eq.~\eqref{e:ODE} using the scaling
ansatz. It is important to note that, in our analysis, we have not dropped any terms either arising from the Weyl-fluid-driving Bianchi identity \eqref{e:bianchi} or the master equation \eqref{e:master}.
We have assumed that the solution in the bulk  depends only on $y H/\xi$ but the form of
the master equation implies that this scaling assumption should be a good one.
We are only testing the robustness of our solution in order to estimate the effect of any new physics which might be important at scales around $r_\cs$ and which is not embodied in the master equation already.

\begin{figure}[ht]\begin{centering}
\includegraphics[width=\columnwidth]{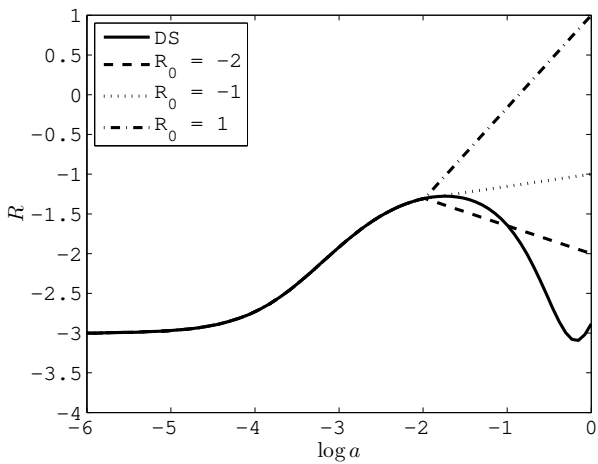}
\caption{\label{f:modR} Off-brane gradient for $k=0.001$~Mpc$^{-1}$
for the scaling solution and the scaling-violating scenarios
employed in robustness testing. Since any scaling violation is only
likely to occur at scale factor close to $Hr_\cs \sim 1$, we modify
the gradient starting at $\log a =-2$. A value of the gradient
today, $R_0$, is chosen and the gradient is interpolated linearly
between these two scale factors. In this modification, we disregard
the fact that the mode enters the horizon at late times.}
\end{centering}\end{figure}

One way of testing the robustness of the scaling ansatz is to change
the values of $R$ at late times and investigate how much of a
departure from scaling-ansatz values is necessary to significantly
change the behavior of observables. We concentrate on the change to
the evolution of the potential $\Phi-\Psi$, which drives the ISW
effect, as we alter the off-brane gradient. Since the QS solution
already has a significantly sharper decay than $\Uplambda$CDM, and
therefore is a worse fit to the large-angle CMB anisotropy
\cite{Song:2006}, and the DS solution decays even more rapidly
(see figure~\ref{f:phi}), we attempt to violate the scaling solution
in such a way as to soften this decay. We present the modification
to $R$ in Fig.~\ref{f:modR}: we employ a linear interpolation for
$R$ between its scaling-ansatz value at $\log a = -2$ and a chosen
off-brane gradient value today, $R_0$. One should note that this
breaking is rather extreme, since the scale under consideration, $k
= 0.001$~Mpc$^{-1}$, is inside the horizon today and it should be
well within the quasi-static regime at the present time.

We have found that choosing negative values for $R_0$ strengthens
the decay of $\Phi-\Psi$, with the scaling solution approximately
replicated for $R_0 \approx -2$. Positive values of $R_0$ reduce the
rapidity of the decay at late times: the QS solution is matched for
$R_0 \approx 5$, which is a value much higher than ever achieved by
the scaling solution. In order to achieve the low levels of decay
exhibited by $\Uplambda$CDM, $R_0$ needs to be set in the vicinity
of 50.

The effect is different for $\Delta$: here, negative values of $R_0$
also decrease the growth rate even further beneath that of DS;
however, $\Delta$ asymptotes to a value approximately 2\% beneath
that of the QS solution as $R_0$ is sent to infinity.

The above considerations show that the new physics required around
$r_\cs$ would have to violate the scaling behavior rather strongly
in order to give an ISW effect comparable to that of $\Uplambda$CDM.

\begin{figure}[ht]\begin{centering}
\includegraphics[width=\columnwidth]{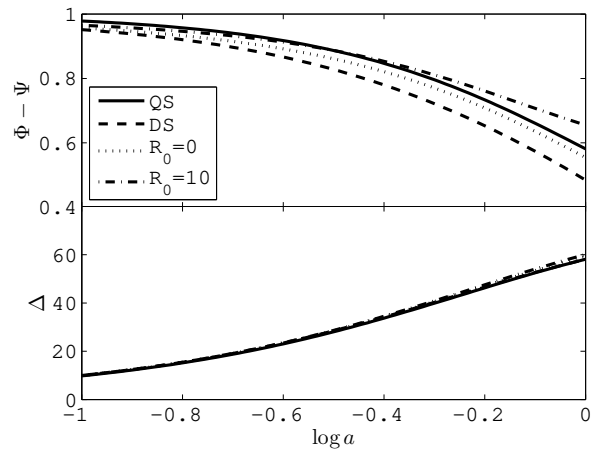}
\caption{Evolution of $\Phi-\Psi$ and $\Delta$ for mode
$k=0.001$~Mpc$^{-1}$ for a selection of scaling-violating scenarios.
Increasing $R_0$ brings the solution closer to that of QS, and, for
very large values, reduces the decay of $\Phi-\Psi$ to that of
$\Uplambda$CDM. The effect of changing $R_0$ on $\Delta$ is much
smaller, the quantity remains insensitive to the precise details of the scenario.}
\end{centering}\end{figure}

\begin{figure}[ht]\begin{centering}
\includegraphics[width=\columnwidth]{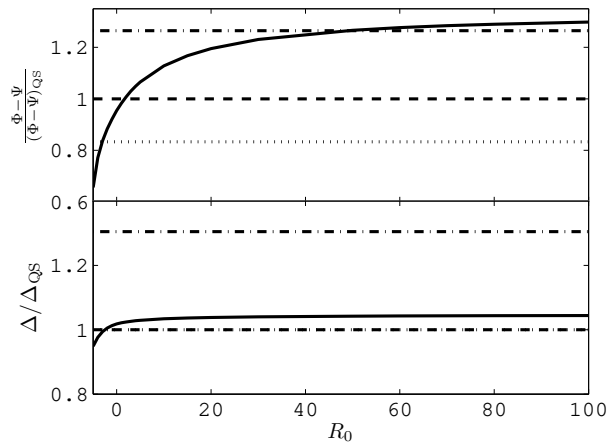}
\caption{Ratio of $\Phi-\Psi$ and $\Delta$ for scaling and
scaling-violating scenarios to their values in QS for mode
$k=0.001$~Mpc$^{-1}$. Dashed line represents the final value of the
quantity in QS, dotted---in DS, and dot-dashed---in $\Uplambda$CDM.
The solution is very sensitive to negative values of $R_0$, but
choosing such scenarios only increases the decay, strengthening the
ISW effect. Positive values of $R_0$ bring the evolution of the
observables closer to that of the QS solution and, for very large
values, achieve decays as low as those of $\Uplambda$CDM. $\Delta$ is quite insensitive to the choice of $R$. All
quantities were normalized to the same value at $\log a = -2$.}
\end{centering}\end{figure}


\clearpage

\end{document}